# Artificial Intelligence enhanced Security Problems in Real-Time Scenario using Blowfish Algorithm

**Yuvaraju Chinnam** [1], **Bosubabu Sambana** [2]

[1] Department of Computer Science and Engineering,
St. Peter's Engineering College (Autonomous), Hyderabad
chinnamyuvraj@gmail.com

[2] Department of Computer Science and Engineering,
Lendi Institute of Engineering and Technology (A),
JNTU GV, Vizianagaram, Andhra Pradesh, India.
[1] bosukalam@gmail.com

**Abstract:** *In a nutshell, "the cloud" refers to a collection of interconnected computing resources made possible by an extensive, real-time communication network like the internet.* Because of its potential to reduce processing costs, the emerging paradigm of cloud computing has recently attracted a large number of academics. The exponential expansion of cloud computing has made the rapid expansion of cloud services very remarkable. Ensuring the security of personal information in today's interconnected world is no easy task. These days, security is really crucial. Models of security that are relevant to cloud computing include confidentiality, authenticity, accessibility, data integrity, and recovery. Using the Hybrid Encryption Algorithm, we cover all the security issues and leaks in cloud infrastructure in this study.

## 1. Introduction

Artificial Intelligence techniques imply that prove an ability to better way to understand current and future technology field requirements specially focused on Commerce and Governance systems based on economy AI plays a key role in part of monitoring the business environments with existing strategies, identifying the customers' requirements, and carrying out the vital techniques without or with negligible human intercessions and involvements. Subsequently, it overcomes any barrier between purchasers marking' requirements and the big impact of quality of Service (QoS).Many corporations are transferring their facts garage to the cloud, however imposing effective security features is a ought to-have earlier than they decide to adopt cloud computing.

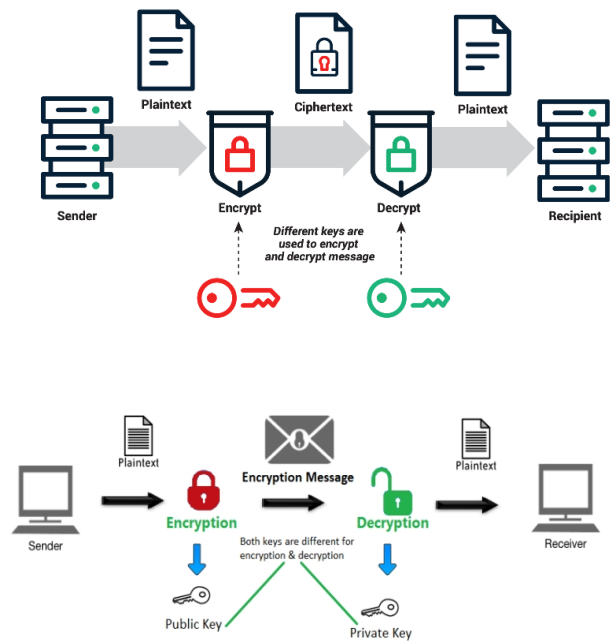

Fighure,1: Crypto Key generation

For facts security in cloud environments, we propose a cryptography symmetric key algorithm for encrypted and decrypted data, the usage of a multilevel cryptography-based safety model [1]. The proposed prototype increases data protection to the most volume feasible for both users and cloud service providers. The model gives the cloud user complete transparency in security.

Nowadays, IT projects are greatly impacted by cloud processing. Businesses are increasingly turning to cloud services because to the many advantages they provide, which are compatible with the rapid pace of technological advancement. Cloud administrations need security and protection on several levels, notwithstanding the many benefits they provide [2].

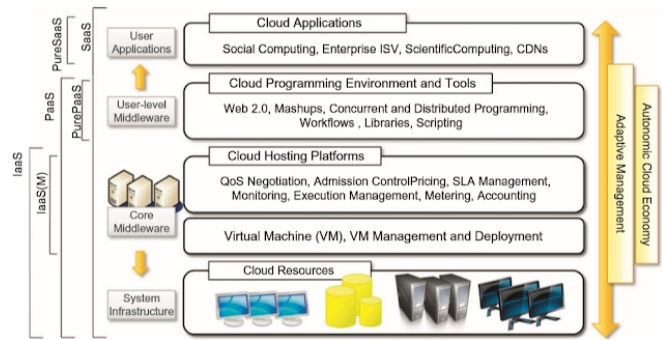

Figure. 2: Cloud Computing Architecture

The increasing sophistication of PDAs has made it possible for users to access cloud services, which in turn allows users to take use of cloud features such as sharing and storing media files across numerous platforms. Information security is always an important aspect of technological advancement. It is imperative that cloud services prioritize security when it comes to vital data that is accessible over the internet. Cloud computing's pervasiveness and the ease with which it may transmit data across borders raise significant security concerns [3]. There are a lot of moving parts when it comes to cloud security, including validation, information security, and protection. For cloud service providers to meet some of these security goals, they must implement certain measures [4].

Data encryption and decryption will be the primary tools for ensuring data security, since privacy is considered a crucial component of information technology. Existing Security methods There are a number of benefits and drawbacks to using algorithms like RSA, Diffie-Hellman, DES, AES, RC4, RC5, RC6, blowfish, W7, and 3 DES for data encryption. According to "Veeraruna Kavitha. (2011), discussed issues facing with security" [5], these algorithms are symmetric and deviated. We want to showcase an encrypted cloud environment that uses both symmetric and asymmetric encryption.

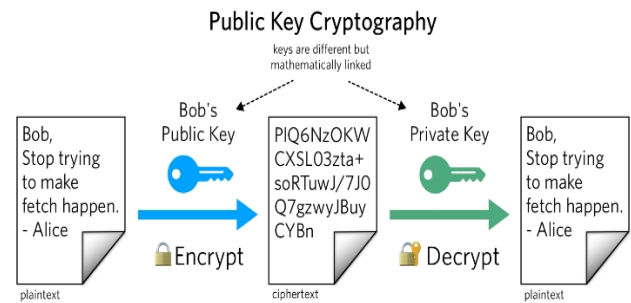

Figure3: Public Key Cryptography

When encrypting and decrypting data, we use RSA asymmetric computation and AES symmetrical calculation [6]. We want to provide a cloud service that ensures security on several levels, including encryption, secret phrase security, multilayered confirmation, and security during the transmission of data [7]. The way client information is protected and processes are safely delegated to a remote cloud service provider become a key issue in Cloud computing.

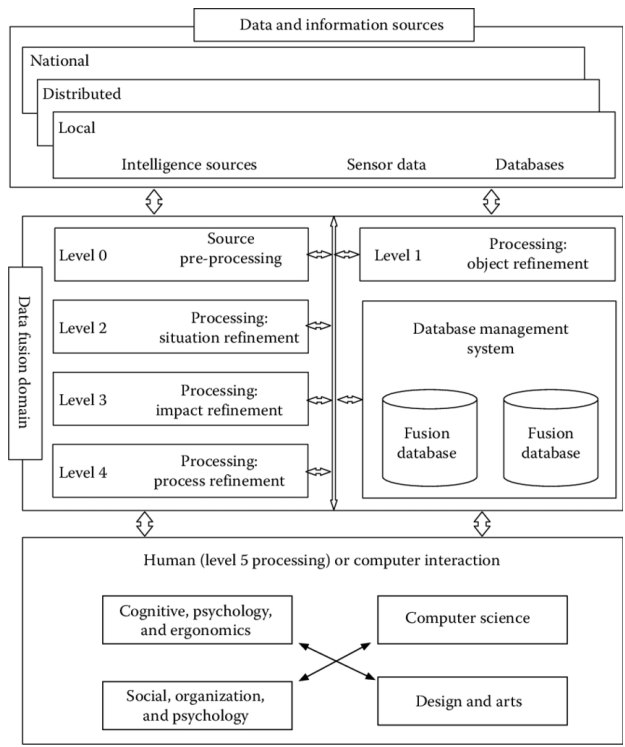

Figure.4: Cloud and data Information sources

Cloud service providers transmit information to clients remotely; therefore, the security of that information is at risk since it can easily be removed by hackers.

To protect the clients' information from hackers, this environment should implement appropriate security measures while operating a variety of harmful activities. Several strategies had been proven to mitigate protection problems, which includes key sharing, the usage of cryptographic algorithms.

Information, its stockpiling, transmission, and usage are analyzed within the context of protection problems and implementation of diverse mechanisms[8].This paper seeks to offer data safety towards special sorts of attackers with the aid of encrypting and decrypting purchaser information at cloud storage and at the purchaser-facet using symmetric key cryptography.

In Inclusion to services of cloud, deployment, securityissues, and constraints in terms of cloud computing. These days, Improving the privacy & securityof cloud data has become a vital concern and the Answer for this is to implement affirmative encryption modules while preserving and pushing the data into the cloud [9].

This entire study puts for wardan innovative hybrid algorithm to improve cloud data security using encryption algorithms.

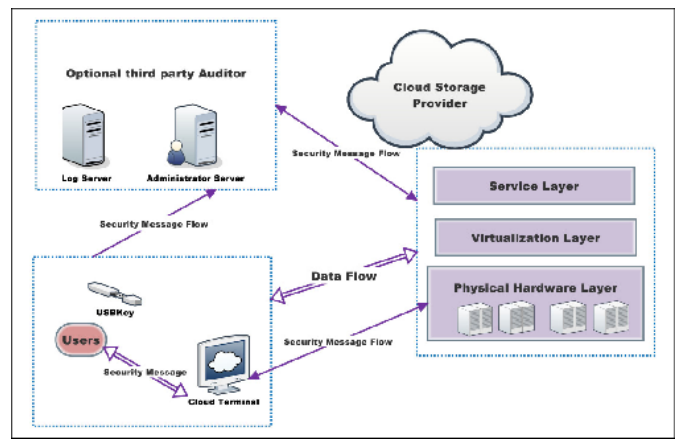

Figure.5: cloud storage architecture diagram

This study syndicates H graphic encryption besides blow-fish toimprove cloud security [10].

Want to conclude that if at all the security problems are solved then theupcoming generations resolve to be the resultsaimed at cloud storage. which could benefit all types of data centers from small-medium to large.

## 2. Literature Review

The security of cloud computing has been the subject of significant research.Several authors have discussed various ways to achieve security, considering that this should be the primary concern for the users.Using Counter Propagation Neural (CPN) networks, A. Negi et al [2 ,28] given a model showing how the encryption and decryption method, can be implemented.

A traditional security system is enhanced with this technology.An information security improvement involves three-tiered authentication. Additionally, the proposed solution incorporates the monitoring of the system in real-time and the operation of the forensic virtual machine. As a result of these techniques, an attacker or unauthorized handlerwould need to decrypt data at every level [11,12], which is a task that is far more challenging or difficult than decrypting it at a single level.

A model proposed by H.S. V et al. [9] can be used to decrease cyber attacks aimed at cloud data confidentiality.Client-side encryption and key management are suggested as a means to improve data security.

A novel cryptographic scheme is created by combining chaos and neural cryptography. The key generation is done using an alphanumeric encryption table.This algorithm is used by users to authenticate themselves, ensuring that data is protected from unauthorized user access.

In [13], Reshma Suryawanshi and her colleagues discuss; primary scheme, public auditing, a homomorphic linear authentication algorithm is used as Third-Party Authentication (TPA) .Meanwhile, their second scheme utilizes threshold cryptography.In the initial scheme, TPA does not learn anything about the important data during the auditing process, whereas the second scheme ensures that unauthorized users cannot misuse the stored data as a result of the audit.

M. K et al.,[14, 15,28] recently proposed a Hybrid Encryption Scheme, it is helpful for the recognizable validation of client or user identity, and through a substantial check, they are for the most part utilized for a biometric interaction.

This proposed Blowfish algorithmic approach design will establish anintensely safe climate and avoid unauthorized or unapproved access. S.Singh et al., [16, 28] proposed a plan that utilizes Elliptic Curve Cryptography. Here, the information is encoded at the customer or user side and must be decrypted done through aftertransferring. Likewise, at the time of login, the client verifies themselves through various information boundaries.

K.Brindha et al. in [17, 28] suggest Visual Cryptography is a system focused on data stockpiling security issues. The data is saved at information servers in an encoded structural formation.

Only subsequently the validation based on authentication of the client would the specialist cooperate through functional data transformation to give keys to the common images. To get the mystery key to the data, the client should superimpose this scrambled picture by this key[18].

In 2011, Jan de Muijnck-Hughes proposed a security technique which is known as Predicate Based Encryption (PBE).PBE represents a family of asymmetric encryption and originates from Identity - Based Encryption [1].

This Predicate Based Encryption focuses its implementation on both Platformsas a service and Software as a service [19]. This proposed technique also precludes unwanted exposure, unwanted leakage, and other unwanted breaches of confidentiality of cloud resident data.

Security Techniques for Protecting Data in the Cloud was a 2011 article by V.S. et.al. With any luck, this article will shed light on the security risks associated with cloud computing and help readers choose the best methods for protecting sensitive data [20].

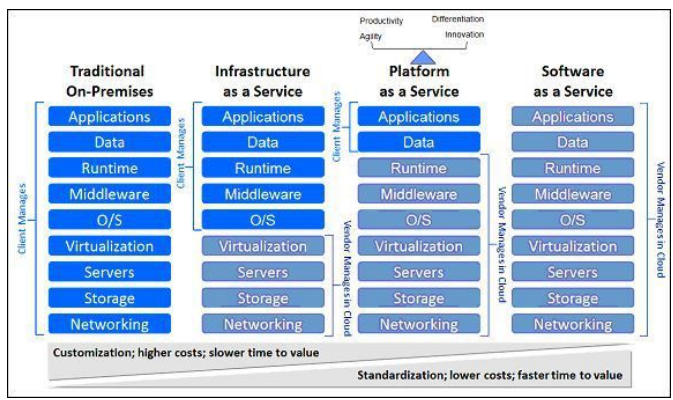

Figure.6.: The Cloud computing Platform-as -a-services Analogy

There are an overall of 43 security threats and 43 countermeasures uncovered by the study [19]. At 31%, confidentiality is the highest assessed feature, followed by availability (19%) and integrity (24%). With the title Security Analysis and Framework of Cloud Computing with Parity Based Partially Distributed File System, Ali Asghar Karahroudy completed a study on the subject in 2011.

A method named Partially was suggested in this article. Client Access Machine, User Public Machine, Cloud Management Server, and PDFSP are the four primary parts of this system [21]. Confidentiality, integrity, and availability are the three pillars of security that this article explored.

Data-Centric Security was a solution-based strategy put out by N. Giweli in 2013. Because this method seeks to provide data-level security, data in the cloud are able to self-describe, self-defend, and self-protect throughout their lifetimes [22]. In this model, the onus for ensuring the confidentiality, integrity, and availability of personal information is squarely on the shoulders of the data owner.

## 3. Related Work

Communication or data transfer over a cloud network must be secure and confidential to maintain the value of the social network. Cryptography takes care of data security. Cryptography manages the security of records that can be saved or transmitted thru the cloud. Secure and private communications or statistics transmission is a need of daily life styles[28].

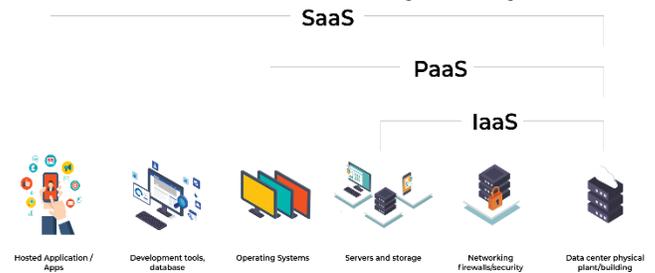

Figure.7: Cloud Platform as Services

Cloud cryptography encrypts data from unauthorized access in the cloud, this makes it possible for users to acquire shared cloud data securely and easily.Using encryption in the cloud, data can be protected from unauthorized access and users can confidently access shared cloud data. Users can utilize cloud encryption to safely access data stored in the cloud.

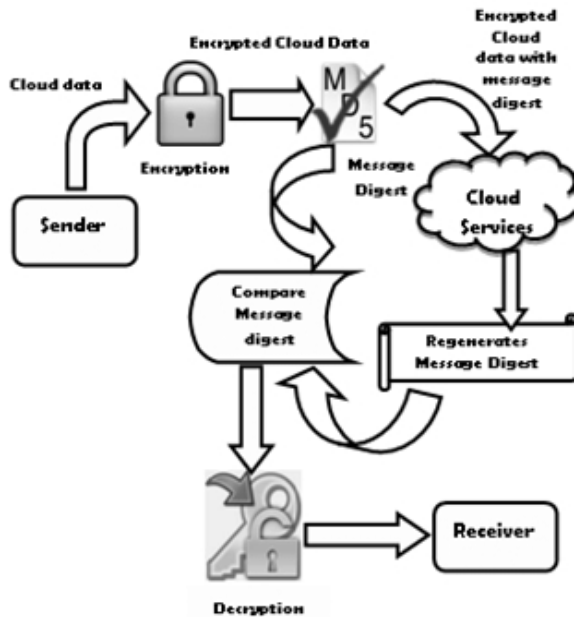

Figure.8: Cloud Platform Cryptography

Cloud providers secure the shared data they host with encryption techniques. Cryptography guarantees the integrity of the information without having to wait for cloud exchange.

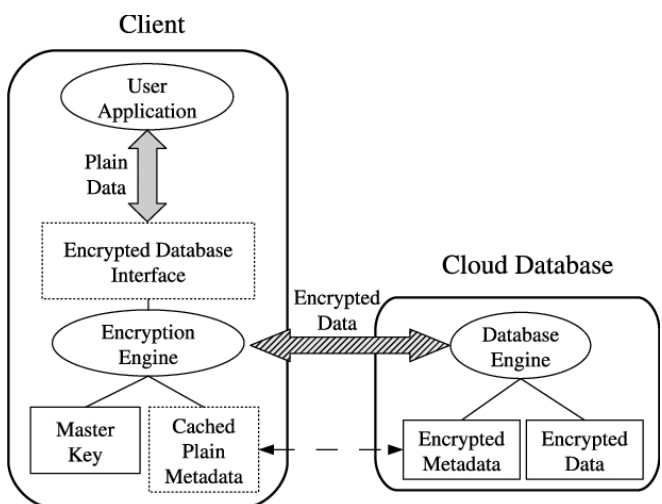

Figure.9: Encrypted Cloud database architecture

Using encryption techniques, cloud providers secure shared data without having to exchange information with each other.Encryption techniques can secure data sharing in the cloud while it is not waiting for data exchange. Cloud service providers are providing shared data and they are securing it with encryption techniques.

The security of data stored in the cloud is ensured by a multitude of algorithms developed using cryptography methods. Advanced cryptographic techniques have become crucial for data protection in the cloud, and they were used to develop security algorithms for cloud computing [28].

**Cryptography algorithm for cloud Security**

The main part for the corporations within the cloud storage version is protection, as they usually shop their information inside the cloud and get admission to the information anywhere and whenever. With cloud computing, companies generally shop their statistics, which is out there anywhere at any time, however statistics security is their predominant problem for cloud computing.

Data transmission and verbal exchange over heterogeneous and related networks are secured with encryption models [6], the encryption algorithm used for cozy records communication is the key for comfy transmission and communication [6]. A secret's to begin with agreed upon with the aid of the interactive mutual customers, and conveying parties and saved in both parties and kept mystery. Presently, the important thing and the encryption set of rules stay used for scrambling the message beyond to sending it starting with one celebration then onto the following [23].

This text received, known as ciphertext, is acquired by way of the other birthday party who then decrypts it taking use of the same key and the decryption set of rules. At this time, the key is maintained as a mystery whilst the encryption and decoding algorithm stays the recognized key additives. As we currently understand, the secret is regarded to both the source and the destination on this cryptography machine, yet this key move is of brilliant importance and ends up being an undeniably hard challenge.

**Data Encryption Standard Algorithm**

Data Encryption Standard is one of the most famous symmetric-key block ciphers published by NIST in 1977.The encryption strategy for DES is exceptionally interesting, as it got contracts a 64-bit plaintext at the source end and produces a 64 ciphertext at the receiverendpoint.

The DES algorithm uses a 48-bit unique key generated by 16 rounds of Feistel structures, even though the key size in DES is 64 bits; despite the fact that the important thing size in DES is 64 bits, simplest fifty four bits are used for encryption and decryption.DES is based totally on Feistel Cipher implementation and makes use of a 16-spherical Feistel structure. This creates an gold standard 48-bit key impartial of the key length, that is in line with the predefined DES algorithm[10,28].

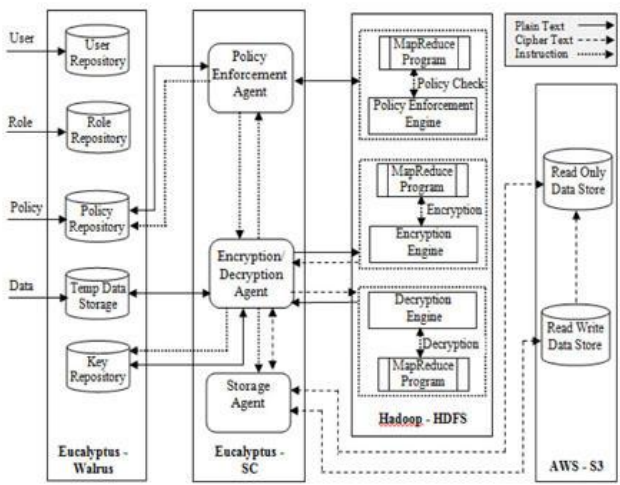

Figure.10: DBMS Implementation on Hybrid Cloud Services

Initially, a 64-bit block of data is permutated, then it is divided into two 32-bit sub-blocks that are presented are subsequently passed into Feistel rounds.Until 16 rounds of encryption are completed, The variety of two folds is elevated and the safety degree is increased, in the ultimate round (after the 16th spherical), the pre-output is generated through swapping L15 and R15 bit portions.

Using both asymmetrical and symmetrical encryption methods, this solution is based on the Chinese Remainder Theorem (CRT). Since the information within the file doesn't need to be encrypted more than once and sophisticated key generation procedures aren't necessary, the suggested approach is shown to be extremely effective in this work [14].

**Algorithm**

Cloud security may be improved using numerous algorithms; however, this part focuses on integrating homographic and Blowfish techniques for optimal cloud security. When it concerns protecting sensitive information, Ahmad and Khandekar (2014) used homomorphic encryption, a novel security technique that allows for the disclosure of assessment results based on encrypted data without revealing the original material used for the evaluation.

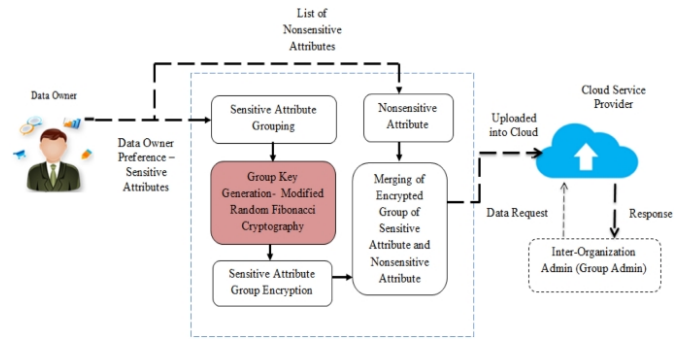

Figure.11: Cloud Service Provider Operation

One of Bruce Scheiner's most comparable publicly available algorithms was the blow-fish algorithm, say Kaur and Kinger (2014). There is currently no known attack that can break the blow-fish cipher, which uses a key of variable length and a 64-bit cipher block. In terms of power consumption and throughput, the Blow-Fish method outperforms other algorithms. According to Rama porkalai (2017), the best algorithm for symmetrical keys is the International Data Encryption Algorithm (IDEA), which requires 64 bits of plain text and a key size of 128 bits. With the exception of one round, all nine of IDE A's rounds are applicable [25]. Cloud computing is the latest trend in IT, according to Gowthami Saranya and Kousalya (2017), although security is the main concern in this area.

A new security measure is developed daily, but it will not provide long-term protection. When it comes to security in a cloud computing setting, encryption, in its many forms, is your best bet [26]. Thus, it can be extrapolated that cloud computing supplies many storages pace to users and security to that data to make the cloud much powerful in the future.

Due to the increasing number of people who use third-party cloud storage services without any authorized permission [27], the security concerns of cloud storage have become the main concern of cloud security. This work presents a symmetric key hybrid multilevel (MLSP) and an asymmetric key algorithm (AKK) that is used to secure data in web-based cloud storage platform services.

Data is moved between data centers is frequently not encrypted, resulting in data loss, private information being shared, government intrusions, intellectual property issues, and snooping efforts.

SSL uses an encrypted link to establish a connection between a web server and browser during data transmission [22,28], but encrypting data before transmission adds a layer of security. Therefore, securing data while in transit is what this algorithm is primarily aimed at. As part of the proposed model, DES and RSA are used for securing the text data while it is being transmitted over the network by using encryption and decryption respectively [26, 28].

The first step for any user who wishes to send a text file is to upload the file to cloud storage, and the second step is for the receiver to download the file from the storage and decrypt it using DES or RSA.

**Proposed Methodology:**

Improving cloud security via the use of encryption methods is detailed in this section. Since data security is a big concern for all users, the research's recommended solution must take precautions to protect users' cloud data. The algorithms used in this research are Blowfish and homographic encryption. Providing the input text is the first stage in this study's use of python software tools and cryptographic methods to improve cloud security.

Cryptography consists of two phases, which are encryption and decryption; both encryption and decryption play a critical role in ensuring data security. Plain text or secret messages are encrypted by using a secret key into ciphertext, a strange message, or a scrambled message. After encryption, the ciphertext is decrypted by using the same secret key used during encoding.

To ensure security, a secret key need to be kept secret since different combinations of the same plaintext result in different ciphertexts.In cryptography, there are numerous different schemes available for encryption, which are used to protect sensitive information. It is a multilayer cryptography algorithm with homographic encryption in the first layer and blowfish encryption in the second layer. The first layer of homographic encryption is applied to the input text.

Then the encryption result will be obtained. After that, the result of encryption is passed to this cold layer which is the blowfish encryption layer. The final output of the encryption layer is obtained.

## Homographic Encryption

If the client is the only one who has access to the secret key to decrypt data using a homographic encryption method, then no one else needs to know the private keys either. The decrypted result of an operation is just as useful as if the computation had been done on the raw data beforehand.

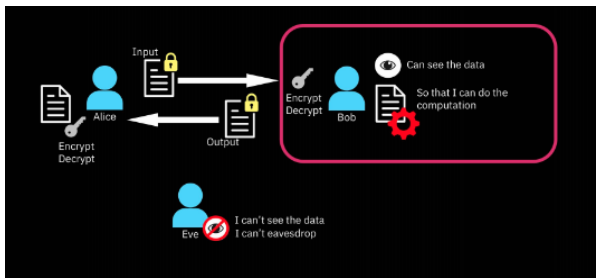

If it is feasible to evaluate the Encrypt function (x, y) from Encrypt (x) and Encrypt (y), where the function might be+ or Ï, and without using secret keys, then the encryption is homographic. Both the Pailler and cryptosystems of Gold wasser-Micali and the ElGamal and RSA cryptosystem are distinguished homographic encryption methods that allow for the evaluation of raw data.

The former uses additive homographic encryption, which merely adds raw data, while the latter uses multiplicative homography.

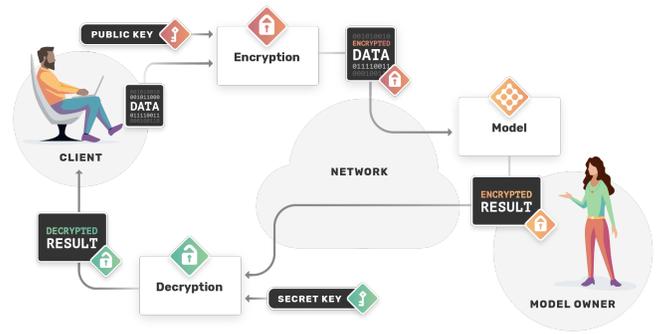

Ae is an algorithm of encryption with key a
Ad is an algorithm decryption

Ad(Ae(m)×Ae(n))=m×n OR
Encrypt(a(×)b)=Encrypt(a)(×)Encrypt(b)
Af(Af(m)×Af(n))=m+n OR
Encrypt(a(+)b)=Encrypt(a)(+)Encrypt(b)

## Blowfish algorithm

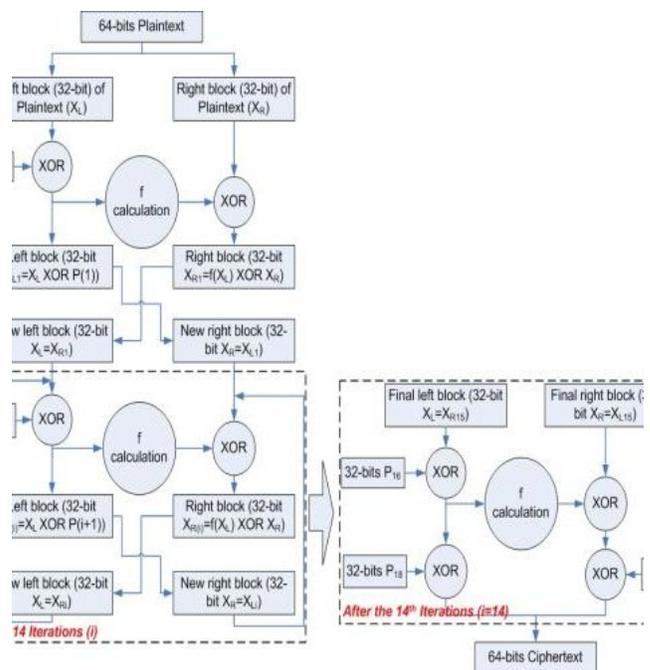

Figure.12: Blowfish Algorithm

Cloud security and privacy issues are created using the Blowfish algorithm. To create the security key, the blowfish algorithm is used. Then, for both methods of encryption and decryption, the asymmetric key block is created. Even inexperienced operators will have trouble accessing the blow-fish key file, even if it is publicly available on the network. Given that the blowfish key technique is among the most secure cipher blocks. The program of cryptography profited from the familiarity with the participation of this study endeavor and stored the file in cloud surroundings securely. If the client needs the case, they may get the material by simply looking for it by name. Follow these steps to get the file you need if their requirements are met. Using the provided properties, the name may be decrypted.

A secret policy associated with a name that, when fulfilled, encrypts random data using the user's characteristics as its key. After the data has been decrypted, the actual file itself can be extracted using the random key. The file has been securely encrypted if unauthorized individuals are unable to access it. The client doesn't consider it acceptable since it cannot decode the content name efficiently, thereby preventing it from accessing the genuine file. Since the user lacks the authority to decrypt the material, they are unable to access it (Saranyaand Kavitha2017).

A network encryption/decryption scheme is proposed in which DES and RSA are used to secure text data during its transmission. For instance, if the user would like to send text files, they must upload them into cloud storage and then apply DES encryption and RSA encryption techniques to the received file [28].

Steps involved in implementing the proposed system

**Sender:**
  **Encryption:**
   a). Uploading the text file by the sender on cloud storage.
   b). At the initial level of encryption, applying the DES and followed by RSA.
   c). At the end, conversion of the plain text into the Cipher Text, maintained in the database.
   d). Foreveryroundr(till18rounds)

  i. XOR left half(L of data with $r^{th}$p-array entry
  ii. Use the
       XORed data as input for f function of the blowfish algorithm
  iii. XOR the F-function output with the right half
     (R)of the data
  iv.   SwapLandRF-Function
  a). Split the 32-bit in put into 4 eight/bit quarters, which are input to s boxes
  b). S-boxes32bitoutput
  c). Outputs are ordered modulo 232 and XORed to create an output of 32 bits and after the 16th spherical XOR L with K16 and R with K17 without the usage of the remaining swap.

**Receiver:**
  **Decryption**
While the steps for encryption and decryption are identical, the sequence in which P1, P2,..., and P18 are used is reversed during decryption. Python is a programming language and package that is used to run the code. Python is a very versatile, high-level, interpreted programming language with a wide range of potential uses.
The principles of Python are:
   a). Simple is good than complex;
   b). Explicit is good than implicit;
   c). counts of readability; and
   d). the complex is good than the complicated.

  **Decryption:**
   a). At the receiver end, read the Cipher Text.
   b). And then, apply the RSA algorithm for the decryption, and after that DES technique for decryption.

The equations of the blowfish algorithm are:
The block size for the algorithm chosen is 64 bits; five sub keys and arrays are used.
  a). 18 entry p array
  b). 256 entries boxes(S0,S1,S2,S3)

The major characteristics of Python are:
  (a). OOparadigm;
  (b). Indentation white space use to represent blocks;
  (c). garbage gathered management of memory;
  (d). dynamic typing;
  (e). interpreted runtime;
  (f). huge third-party libraries repository; and
  (g). huge standard library.

Python is a popular programming language that many companies utilize for many purposes, including but not limited to web development, embedded application development, scientific computing, the development of software, artificial intelligence, and information security.

Figure.13: Process of text conversion at sender and Receiver side

## 4. Conclusion

This paper proposes secure and reliable encryption and decryption algorithms for cloud storage. It enhances the security of data in the cloud environment. Security is the issue when it comes to storing information in the cloud. This paper recommends a multilevel encryption algorithm that is used to secure all the sensitive data stored in the cloud.

Faster creativity, more flexible resources, and economies of scale are all possible with cloud computing, which is the delivery of computing services over the internet, servers, storage, databases, networking, software, analytics, and intelligence. People are increasingly making use of cloud computing, which is a relatively new societal phenomenon. When bringing in new technology, there are always going to be problems that make it hard to use. In today's world, cloud computing is considered a rapidly expanding field that may provide an instantaneously expandable service over the internet employing both hardware and software virtualization.

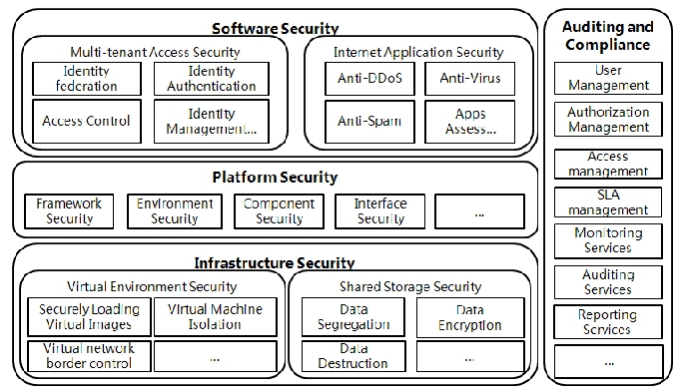

Figure .14: Data Security in Cloud Computing

It is critical to pay close attention to the issue of data security in cloud computing. So, it's crucial to do thorough research on how to design and execute effective security measures that will keep hackers out of cloud data transmissions. From what we can tell from this study's literature review and practical application of the suggested methods, it seems that while most papers focus on data confidentiality, very few manage to meet all three requirements of security: availability, integrity, and confidentiality.